\documentclass[12pt,preprint]{aastex}

\shorttitle{$\gamma$-ray observations in LIRGs.}
\shortauthors{Cillis et al.}

\received{TBD}
\begin{document}

\title{EGRET upper limits and stacking searches of $\gamma$-ray observations
of luminous and ultra-luminous infrared galaxies}

\author{Anal\'{\i}a N. Cillis\altaffilmark{1,2}, Diego F. Torres\altaffilmark{3}, \& Olaf Reimer\altaffilmark{4}}

\altaffiltext{1}{Code 661, LHEA NASA Goddard Space Flight Center,
Greenbelt, MD 20771, E-mail: cillis@blazar.gsfc.nasa.gov}
\altaffiltext{2}{Also SP Systems Inc.}
\altaffiltext{3}{Lawrence Livermore National Laboratory, 7000 East
Ave., L-413, Livermore, CA 94550, E-mail: dtorres@igpp.ucllnl.org}
\altaffiltext{4}{Institut f\"ur Theoretische Physik IV,
Ruhr-Universit\"at Bochum, 44780, Germany. E-mail:
olr@tp4.ruhr-uni-bochum.de}

\begin{abstract}

We present a stacking analysis of EGRET $\gamma$-ray observations at
the positions of luminous and ultraluminous infrared galaxies. The
latter were selected from  the recently presented HCN survey, which
is thought to contain the most active star forming regions of the
universe. Different sorting criteria are used and, whereas no
positive collective detection of $\gamma$-ray emission from these
objects we determined both collective and individual upper limits.
The upper most excess we find appears in the case of ULIRGs ordered
by redshift, at a value of 1.8$\sigma$.

\end{abstract}


\section{Introduction}

Luminous infrared galaxies (LIRGs) are the dominant population of
extragalactic objects in the local ($z<0.3$) universe  at bolometric
luminosities above $L > 10^{11}$ L$_\odot$. Ultra luminous infrared
galaxies (ULIRGs) present $L_{\rm FIR} > 10^{12}$ L$_\odot$, and are
the most luminous local objects (see Sanders \& Mirabel 1996 for a
review). The most notable feature of LIRGs and ULIRGs is perhaps the
large concentration of molecular gas  that they have in their
centers, at densities orders of magnitude larger than found in
Galactic giant molecular clouds (e.g., Downes et al.  1993; Downes
\& Solomon 1998; Bryant \& Scoville 1999). This large molecular
material is believed to be the result of a merging process between
two or more galaxies, in which much of the gas in the former spiral
disks
---particularly that located at distances less than 5 kpc from each
of the pre-merger nuclei--- has fallen together, therein triggering
a huge starburst phenomenon (e.g., Sanders et al. 1988; Melnick \&
Mirabel 1990). LIRGs and ULIRGs thus have large CO luminosities and
a high value for the ratio $L_{\rm FIR}/L_{\rm CO}$, both being
about one order of magnitude greater than for normal spirals. The
latter substantiates, based on star formation models, a greater star
formation rate per unit mass of gas.

In a recent letter \cite{tor2004},  by computing the
$\gamma$-ray flux produced by the interaction between an enhanced
cosmic ray population and the molecular material, ultimately leading
to neutral pion decay, it was shown that LIRGs and ULIRGs are
plausible sources for GLAST and the next generation of Cherenkov
telescopes. This result was deepened by a detailed analysis of the
$\gamma$-ray emission from Arp 220,  the most extensively studied
ULIRG, that included the emission of secondaries \cite{torres2004}. The
enhanced population of relativistic particles is the result of the
large number of supernovae and young stellar objects present in the
central environment. The star formation rate in LIRGs is 100--1000
times larger (e.g., Gao \& Solomon 2004b; Pasquali et al. 2003) and
scales with the amount of dense molecular
gas (traced in turn by the HCN line). It is natural to expect that
the central regions of LIGs have cosmic ray enhancements comparable
to the ratio between their star formation rates (SFRs) and that of
the Milky Way. Torres et al. (2004) showed that if this is correct,
then many of the LIRGs --particularly those located in the 100 Mpc
sphere-- are going to appear as individual GLAST sources.

Although detailed predictions for a particular ULIRG (Arp 220,
Torres 2004) suggest that these galaxies were below EGRET
sensitivity, it was yet open to discussion if they would show up in
an stacking search. This search, similar to what was made by Cillis
et al. (2004) for radiogalaxies, and by Reimer et al. (2003) for
clusters of galaxies, is presented here. We also provide here upper
limits from existing EGRET data for the fluxes of LIRGs in different
energy bands, which are useful both, for future theoretical
modelling and for consistency check with new sets of data.

\section{Stacking technique}

The general stacking method we have applied follows that outlined by
Cillis et al. (2004) when studying radiogalaxies. In order to
perform the stacking technique and look for a possible collective
detection of the $\gamma$-ray emission from LIRGs, we have extracted
rectangular sky maps with the selected target objects located at the
center.

We have used EGRET data from April 1991 through September 1995 ---as
covered by the Third EGRET Catalog (Hartman et al. 1999), in
celestial and galactic coordinates. The extracted maps for each
particular target were chosen to be $60^{\circ} \times 60^{\circ}$
in size, in order to have large off fields of views and be
consistent the EGRET point spread function (PSF).

Transforming the maps to an equatorial position before co-adding
them causes a substantial image distortion, except for those
originally near the all-sky map equator. To minimize this distortion
we have extracted maps from the all-sky map, celestial or Galactic, which
had the target object closer to its equator. This was done for both, count
and the exposure maps, and for all targets. We have transformed the
coordinates of each map into pseudo-coordinates, with the target
object at the center. After doing this, the maps were co-added,
producing the stacking.

It was also necessary to extract a diffuse background map for each
target object.  For this purpose, we have used the diffuse model
that is standard in EGRET analysis \cite{sdh1997}. In order to take
into account the existence of known EGRET sources, idealized sources
with the appropriate fluxes distributed following EGRET's PSF were
added to the diffuse map. This was done only for the sources that
were detected significantly during the time interval of the all-sky
maps: 3EG sources that were significantly detected only during
shorter sub-intervals were not considered for the background model.
It was necessary to normalize each one of the extracted diffuse maps ($D_i$)
for the different exposures ($\epsilon_i$) of the target objects.
The extracted diffuse map for each target object was also transformed
into pseudo-coordinates.
Finally the diffuse maps for the co-added data were obtained as:
$\frac{1}{\epsilon_{tot}}{\sum_{i} c_{i}}$ where $c_{i}$ are the counts
diffuse maps ($c_i=\epsilon_{i} D_{i}$) and
$\epsilon_{tot}=\sum_{i}\epsilon_{i}$.

Stacked maps for the  different groups of objects analyzed below
were created and results are described in the next section. To find
the significance of the detection in a particular ``class'' we have
used the EGRET likelihood ratio  \cite{jrm1996}, a formalism that
produces a ``test statistic'' (``TS'') $TS=-2(\ln L_0 - \ln L_1)$,
where ${ L_0}$ and ${ L_1}$ are likelihood values with and without a
possible source.  $\sqrt{TS}$ is roughly equivalent to the number of
standard deviations ($\sigma$).

\section{Classes of LIRGS and results}

We have stacked galaxies from the HCN survey (Gao \& Solomon 2004a;
Gao \& Solomon 2004b) after sorting them using different criteria.
The HCN survey is a systematic observation   of 53 IR-bright
galaxies, including 20 LIRGs with $L_{\rm FIR}>10^{11}$L$_\odot$, 7
with $L_{\rm FIR}>10^{12}$L$_\odot$, and more than a dozen of the
nearest normal spiral galaxies. Essentially, all galaxies with
strong CO and IR emission were chosen for survey observations. It
also includes a literature compilation of data for another dozen
IR-bright  objects. This is the largest and most sensitive HCN
survey of galaxies, and thus of dense interstellar mass residing
there,  to date.

We have excluded from our consideration those  galaxies that are
close to the Galactic Plane $|b|<20^{\circ}$ (outer Galaxy), and
$|b|<30^{\circ}$ ($|l|<50^{\circ}$), because the background around
those objects would overwhelm any possible signal.

Table 1 shows all the galaxies in the HCN survey referred above,
ordered by $L_{HCN}/L_{CO}$  indicating their coordinates, redshift,
the ratio between line luminosities --$L_{HCN}$ and $L_{CO}$--, and
cosmic ray enhancement. In that Table, the column ``Galactic Plane''
shows whether the galaxy is within $|b|<20^{\circ}$ (outer Galaxy),
and $|b|<30^{\circ}$ ($|l|<50^{\circ}$) and thus whether that galaxy
was excluded from our tests. Following Torres et al. (2004), to
which we refer for details, we have computed the minimum average
value of cosmic ray enhancement, dubbed $k$, for which the
$\gamma$-ray flux above 100 MeV would be above 2.4 $\times 10^{-9}$
photons cm$^{-2}$ s$^{-1}$. The latter is approximately the GLAST
satellite sensitivity after 1 yr of all-sky survey. $k$-values of at
least a few hundreds are deemed probable, based on enhancements
derived in individual supernova remnants.  Luminosity distances used
were those provided in the HCN survey, assuming a Hubble parameter
of $H_0$=75 km s$^{-1}$ Mpc$^{-1}$; although, since redshifts are
very small, changes in the cosmological model do not introduce
significant changes in distances. We selected the galaxies to
perform the stacking technique taking into account the brightest,
the nearest, those having the smaller cosmic ray enhancement needed
to produce fluxes above GLAST sensitivity, and finally the ratios
between the $L_{HCN}$ and $L_{CO}$, and between SFR and $k$. In the
case of the latter, those galaxies having $(SFR/SFR_{MW})/k> 1 $ are
believed to be particular good candidates for detection.

For each class or subclass we have generated stacked maps containing
$N$ galaxies, with $N$=2, 4, 6, etc. For each stacked map so
generated, we have then determined the detection significance using
the standard likelihood method. The results of this research are
summarized in Table 2. For each sorting criterion we have specified
the total number of the objects considered, the maximum detection
significance, and the number of objects yielding that maximum. We
have found no significant result above $1.8\sigma$, for any class
investigated and for any number of objects included. This number was
obtained for the case in which all ULIRGs with $L>10^{12} L_{\odot}$
were considered ordered by redshift.

In Figure 1 we show an example of the stacked maps created (maps of
counts, exposure and background). The left column corresponds to the
case of the highest TS obtained (the case described in the last
paragraph). The middle and right column show  stacked maps for LIRGs
ordered by $L_{HCN}/L_{CO}$ for 54 stacked galaxies (where the TS
obtained was equal to zero), and for the first four LIRGs ordered by
$L_{HCN}/L_{CO}$ (ARP 193, MRK 273, MRK 231, UGC 05101), where the
maximum TS for the class was obtained ($\sqrt{TS}=1.2$).

\begin{figure}
\figurenum{1}
\epsscale{0.30}
\plotone{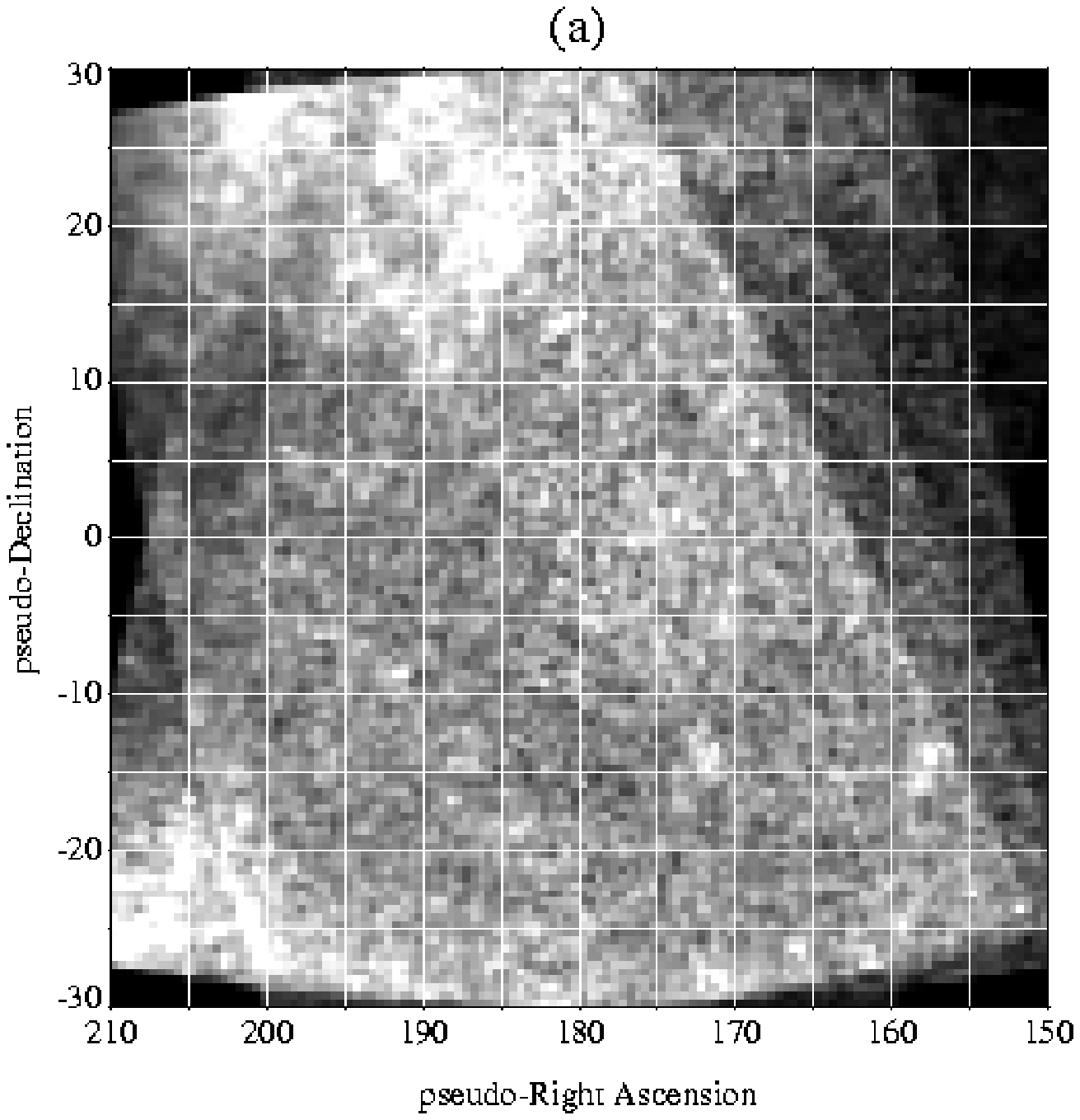} \epsscale{0.30}
\plotone{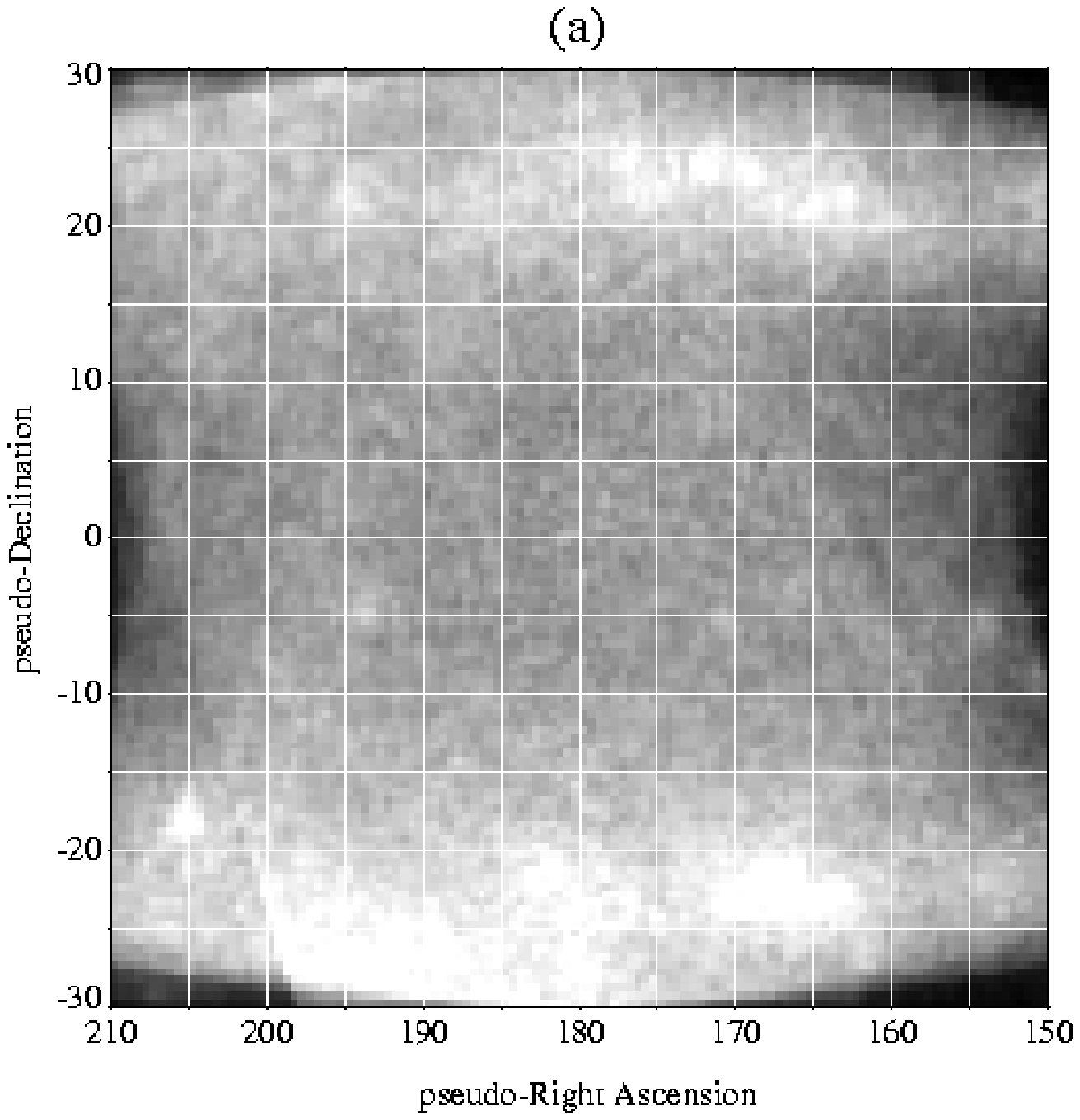} \epsscale{0.30}
\plotone{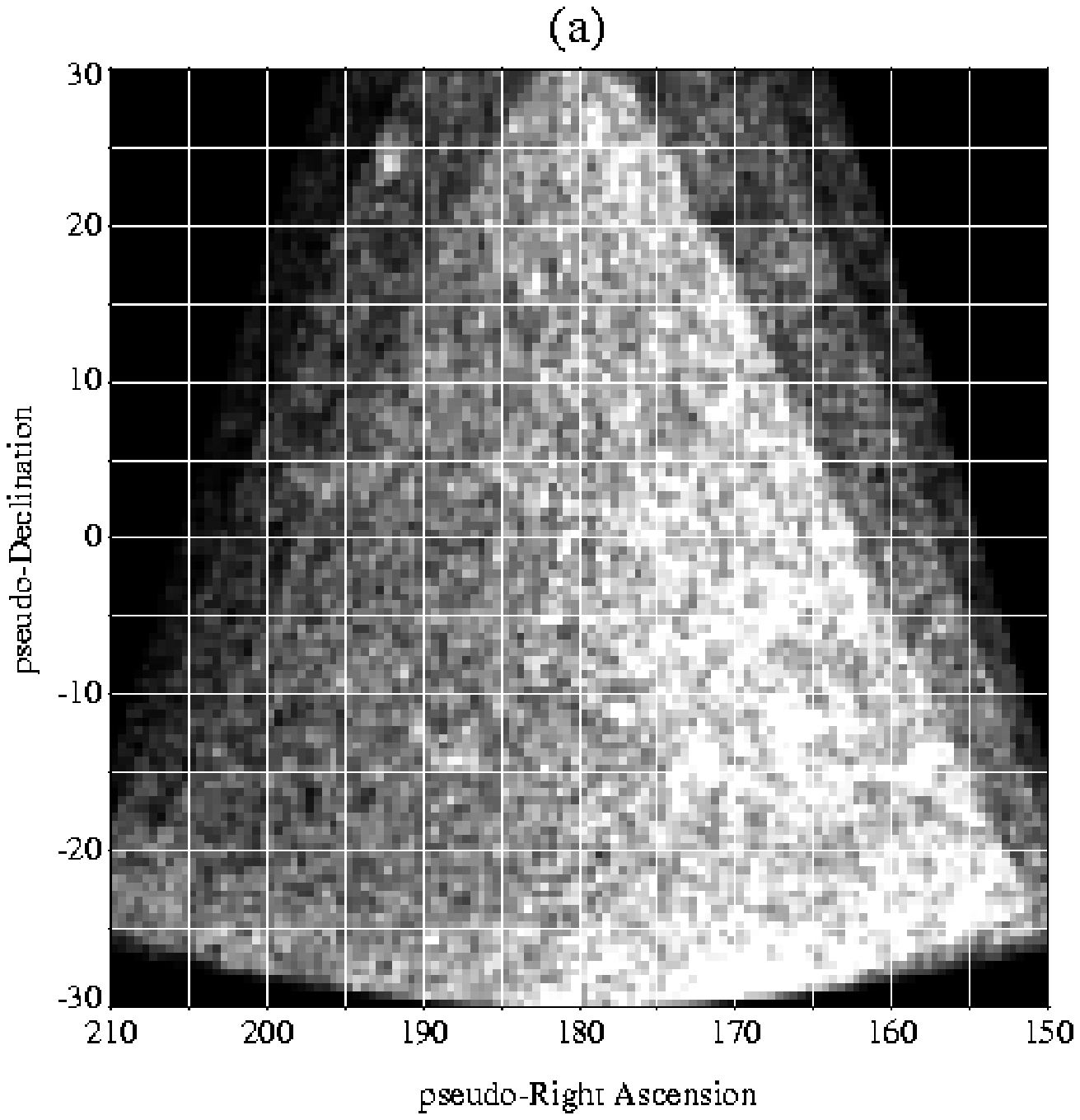} \epsscale{0.30}
\plotone{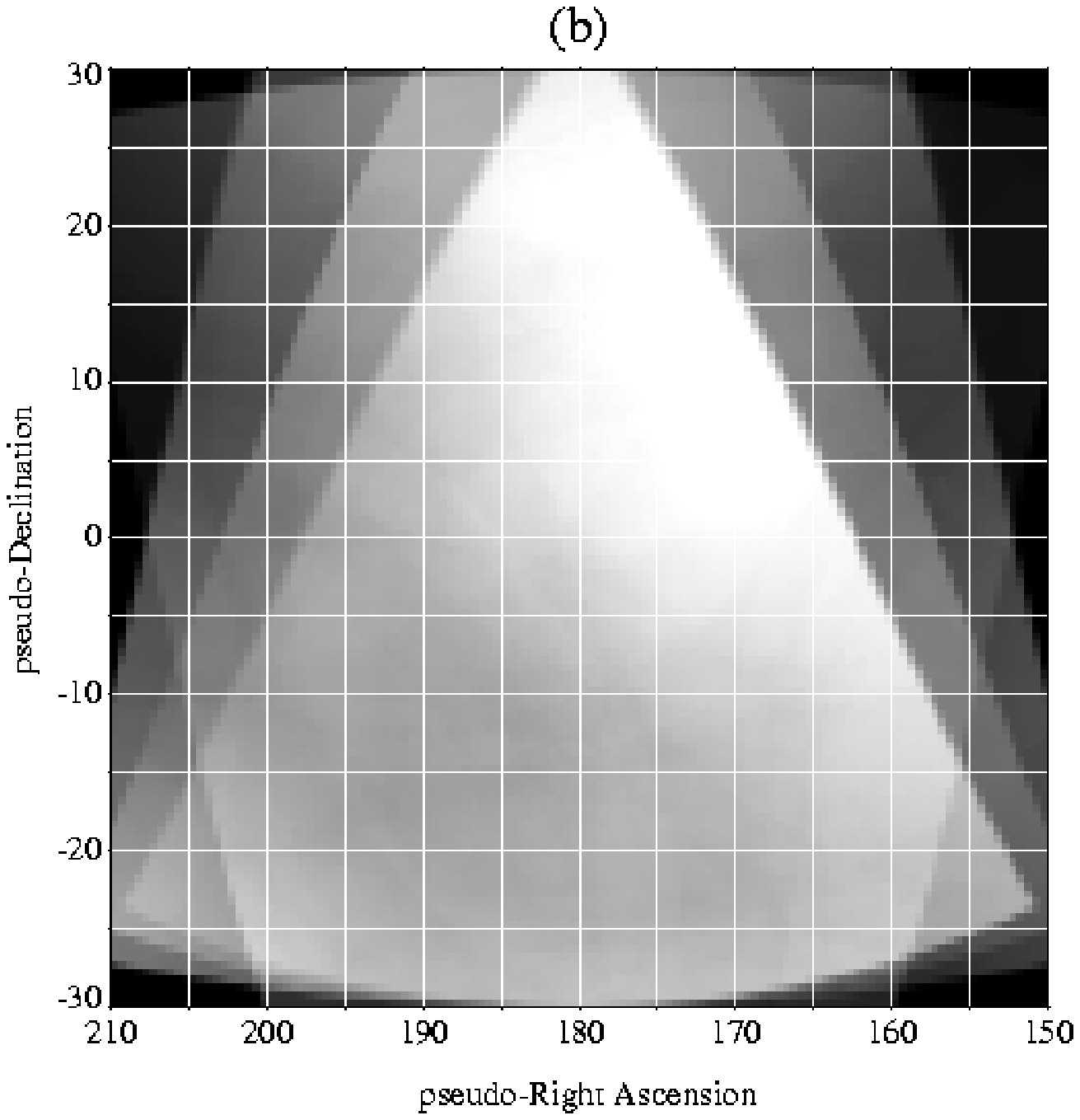} \epsscale{0.30}
\plotone{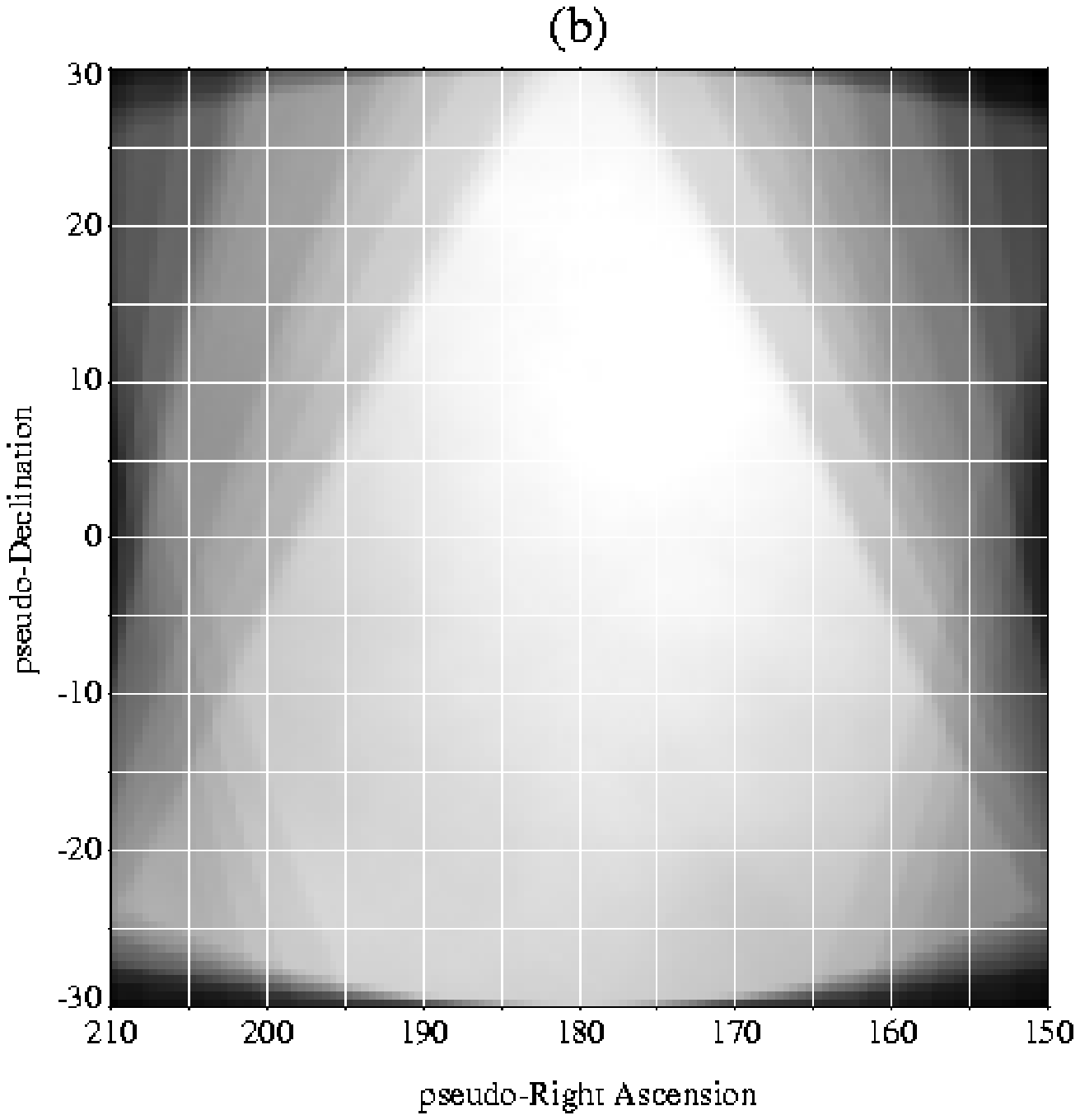} \epsscale{0.30}
\plotone{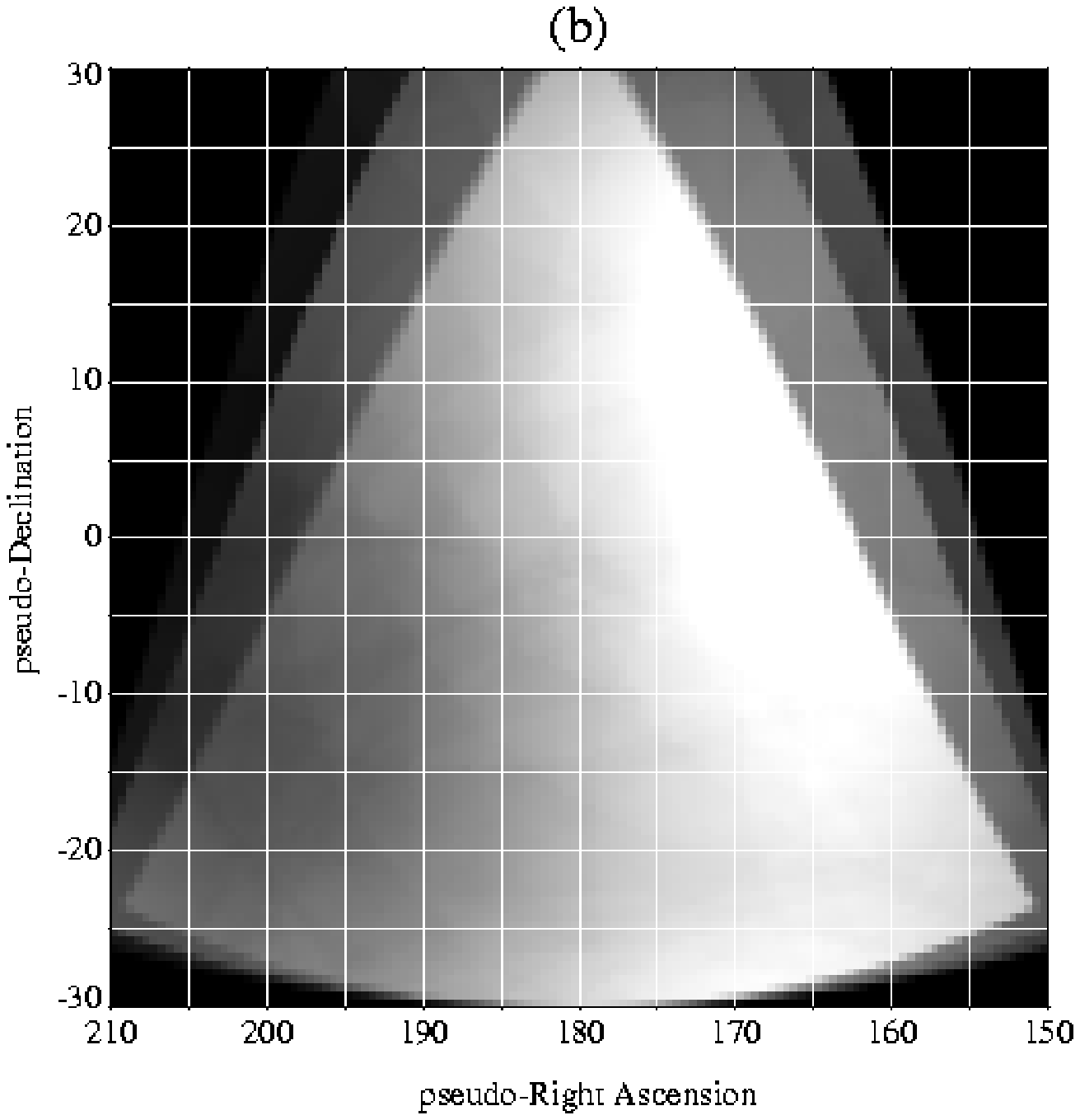} \epsscale{0.30}
\plotone{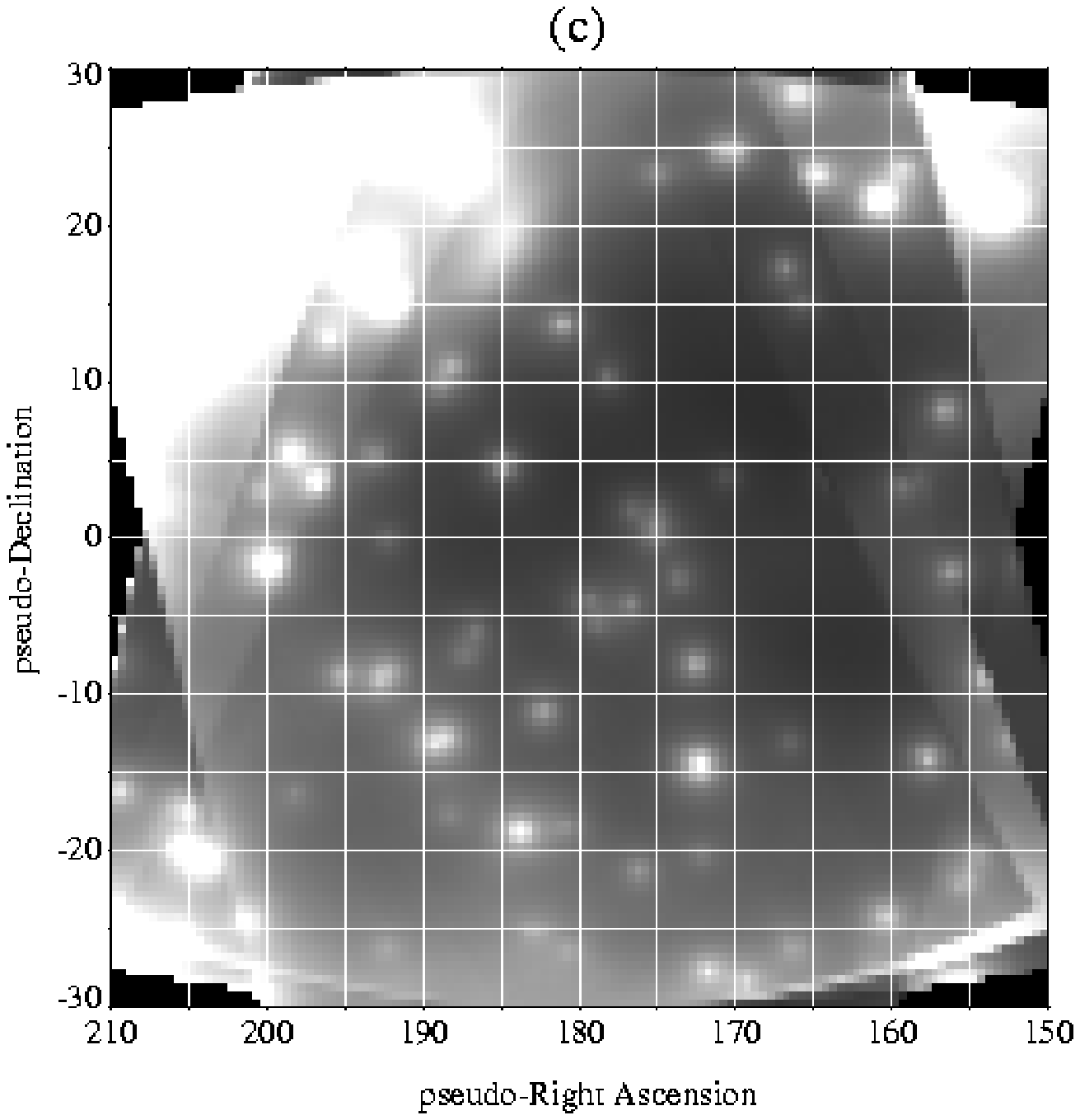} \epsscale{0.30}
\plotone{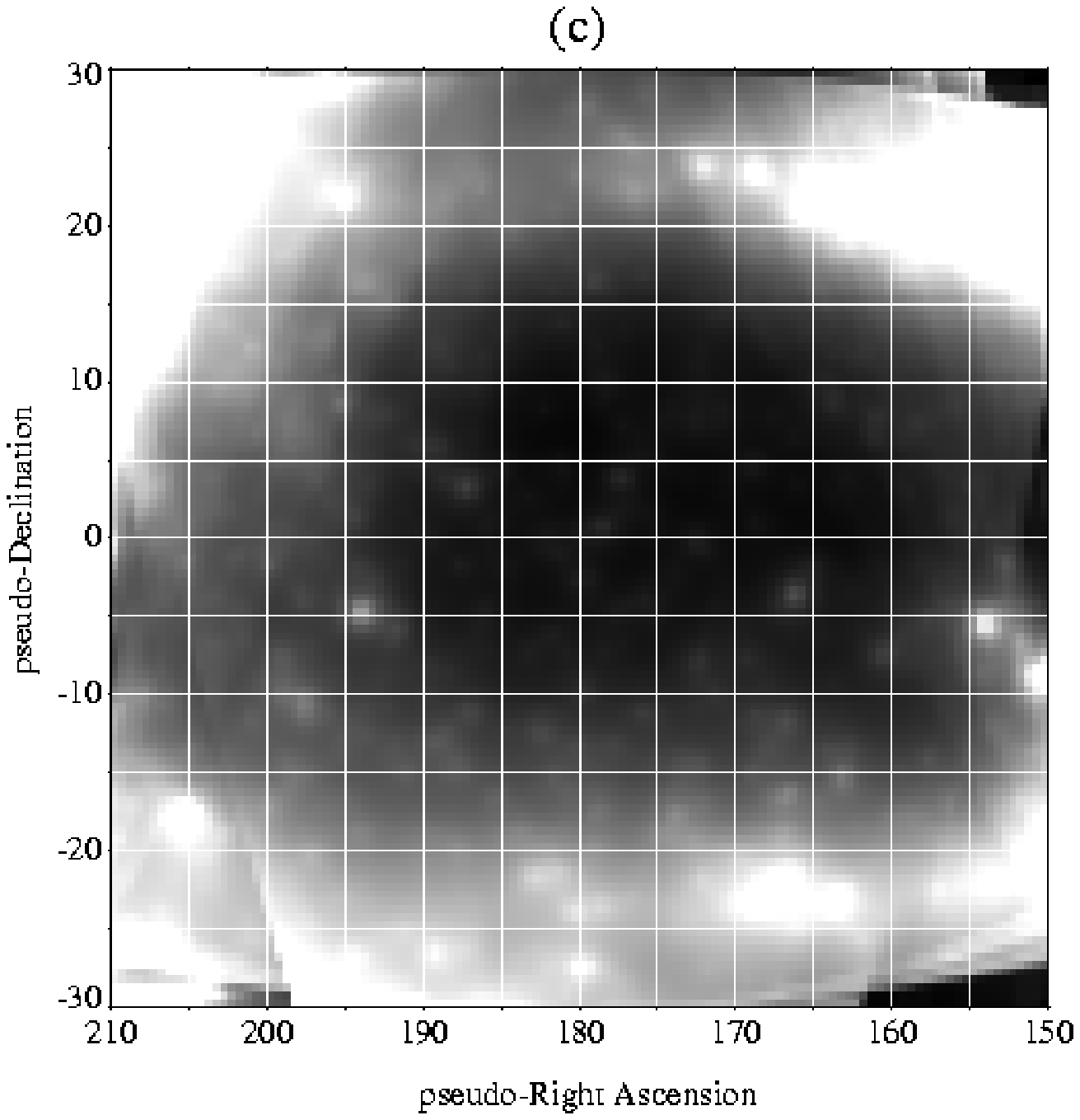} \epsscale{0.30}
\plotone{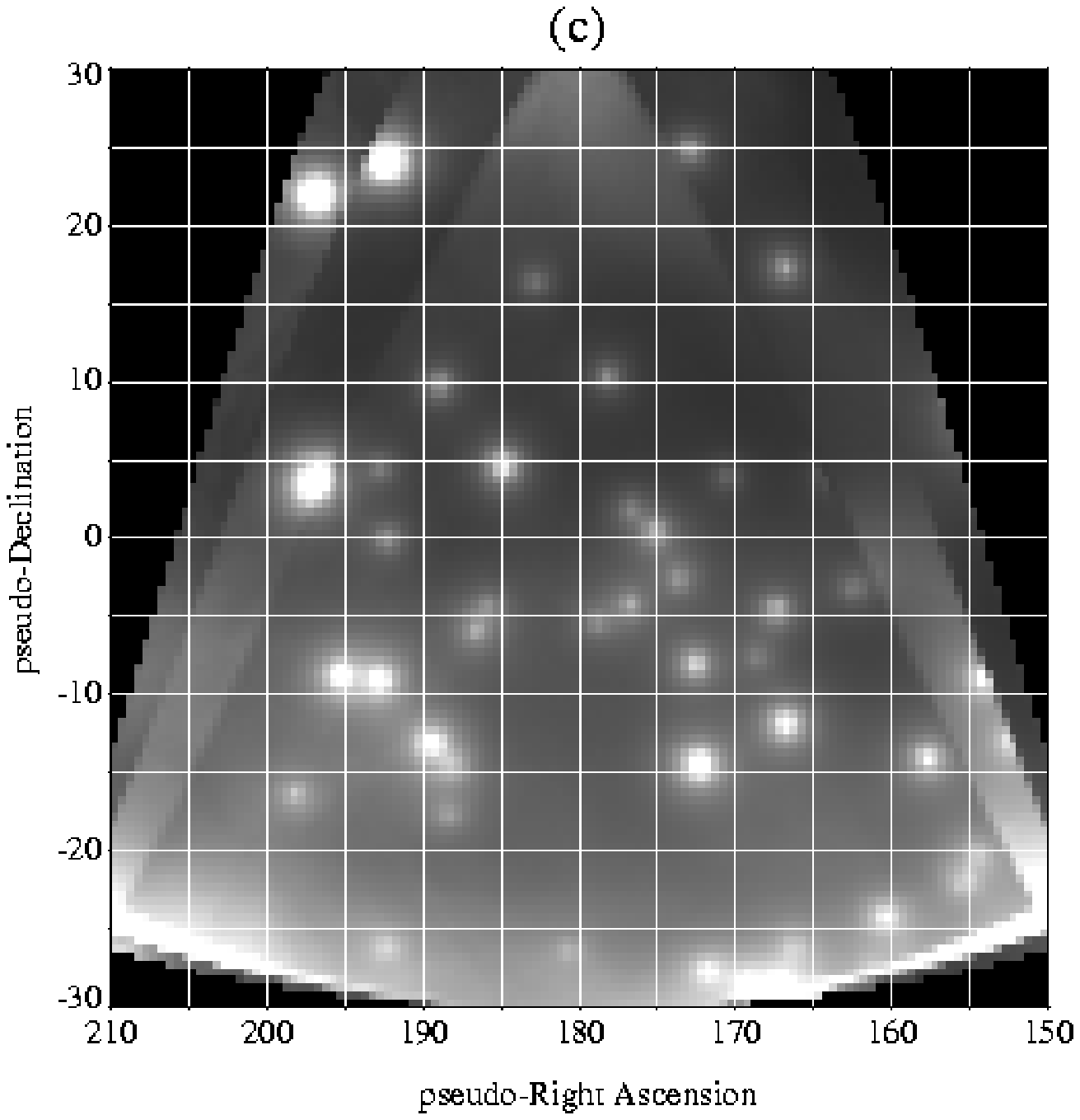}
\caption{First column: Example of
stacked maps for 6 ULIRGs with $L_{IR} \ge 10^{12}L_{\odot}$ ordered
by redshift (ARP220, MRK 273, UGC 05101, MRK 231, 05189-2524,
10566+2448). Middle column:  Example of stacked maps for 54 LIRGs
ordered by $L_{HCN}/L_{CO}$ (see Table 1). Right column: Example of
stacked maps for 4 LIRGs ordered by $L_{HCN}/L_{CO}$ (ARP193,
MRK273, MRK231, UGC 05101). In all columns, (a) counts, (b) exposure
(c) background. The location of interest is at pseudo-Dec $0^{o}$
and pseudo-RA $180^{o}$.}
\end{figure}

Upper limits for the fluxes  for all LIRGs in the HCN survey that
are located away from the Galactic plane are given in Table 3, in
units of $10^{-8}$ cm$^{-2}$ s$^{-1}$. Table 3 presents five of the
sixteen energy bins where we have conducted this
analysis.\footnote{These are  30-50, 50-70, 70-100,100-150, 150-300,
300-500, 500-1000, 1000-2000, 2000-4000, 4000-10000, 30-100,
100-300, 300-1000, $>100$, $>300$, $>1000$ MeV.}  None of the LIRGs
we investigated have been individually detected, which is, in fact,
consistent with the level of flux expected from LIRGs (i.e., fluxes
above GLAST sensitivity but below the EGRET one (see, e.g., Torres
2004). The only galaxy for which a flux (not an upper limit) was
determined is Arp 55. The flux for this galaxy, in the range 500
MeV--1 GeV was $1.9\pm 0.6 \times 10^{-8}$ photons cm$^{-2}$
s$^{-1}$, with $TS=26.3$. This might be thought of as suggestive for
a detection, although Arp 55 was not significantly detected in
consequent energy intervals, and seems not to be special otherwise.
Its redshift is larger than the prototypical ULIRG, and more active
galaxy, Arp 220, and moreover, its $L_{\rm HCN}/L_{\rm CO}$ ratio is
smaller than that obtained for the former. Lastly, one expects
statistical fluctuations when investigating a sample of more than 50
galaxies in 16 different energy bins.

\section{Discussion}

We have presented an stacking search for $\gamma$-ray emission from
LIRGs and ULIRGs  using data from the EGRET experiment. Our results
show that these galaxies were neither individually nor collectively
detected, under a variety of different sub-sampling and ranking
ordering. Apart from the obvious arrangement by distance (redshifts)
we have essentially explored all possible ordering parameters to
investigate the preferable cases for detectability at high energy
$\gamma$-rays. They included the ratio between line luminosities of
HCN and CO; i.e., the ratio between the dense mass most plausible
subject to higher enhancements of cosmic rays and the molecular mass
less densely distributed as traced by CO. They also included the
value of cosmic ray enhancement
---computed under simplifying assumptions-- that would make the galaxy
detectable by the GLAST-LAT. Also, we considered the ordering using
the ratio between the SFR in Milky Way units and the cosmic-ray
enhancement, in the understanding that a realistic value for the
latter should exceed the SFR. We have also imposed upper limits in
different energy bands that can be used as a constraint for future
multifrequency modelling. Even if we presently were not able to
detect LIRGs and ULIRGs in high energy $\gamma$-rays, a suggestive
excess has been found among the most active star forming regions of
the sample, especially when ordered by redshift. Summarizing, LIRGs
and ULIRGs, whereas not individually detected by EGRET (a result
consistent with theoretical expectations) well remain a plausible
candidate for GLAST and Cherenkov telescopes detections.

\acknowledgments

ANC would like to thank R.C. Hartman and D. L. Bertsch for useful
discussions. The work of DFT was performed under the auspices of the
U.S. D.O.E. (NNSA), by the University of California Lawrence
Livermore National Laboratory under contract No. W-7405-Eng-48. OR
acknowledges support by DLR QV0002.

\clearpage


\singlespace

\begin{deluxetable}{llllllllllll}

\tabletypesize{\scriptsize}

\rotate

\tablewidth{0pt}

\tablenum{1}

\tablecaption{HCN Survey}

\tablehead{

\colhead{NAME} &\colhead{l} & \colhead{b} & \colhead{G-Plane} & \colhead{RA} & \colhead{DEC} & \colhead{z} & \colhead{Mag} & \colhead{$L_{HCN}/L_{CO}$} & \colhead{$SFR/SFR_{MW}$} & \colhead{$k$} & \colhead{$\frac{SFR/SFR_{MW}}{k}$}

}

\startdata


\bf{*17208-0014}&22.221135&19.353624&YES&260.841382&-0.283436&0.04281&15.1&0.256&1041.23077&427&2.438479555\\

ARP 193&82.928889&80.5998083&NO&200.1472625&34.1394972&0.02335&14.4&0.238&263.07692&452&0.582028584\\

\bf{*MRK 273}&108.1059064&59.6818904&NO&206.1754626&55.8868475&0.03778&15.07&0.234&420.92308&746&0.564240054\\

\bf{MRK 231}&121.6108323&60.2423486&NO&194.0593079&56.8736769&0.04217&14.41&0.226&515.07692&738&0.697936206\\

\bf{*UGC 05101}&152.476413&42.896978&NO&143.965212&61.353137&0.03939&15.2&0.197&276.92308&1057&0.261989669\\

\bf{*23365+3604}&106.99783&-24.268484&NO&354.755292&36.352417&0.06448&16.3&0.176&415.38462&1743&0.238315904\\

\bf{NGC 1068}&172.1037384&-51.9337526&NO&40.6696292&-0.0132806&0.00379&9.61&0.174&99.96923&28&3.570329643\\

\bf{*10566+2448}&212.455124&64.688165&NO&164.825577&24.542855&0.0431&15.7&0.166&282.46154&1022&0.276381155\\

\bf{NGC 6921}&66.983282&-7.558412&YES&307.12025&25.723417&0.01429&14.4&0.16&77.81538&435&0.178885931\\

NGC 6240&20.728994&27.291006&YES&253.245375&2.400944&0.02448&13.8&0.139&304.61538&255&1.194570118\\

\bf{IC 5179}&6.502243&-55.927177&NO&334.037917&-36.843722&0.01141&12.38&0.129&94.70769&169&0.560400533\\

\bf{*ARP 220}&36.627128&53.02893&NO&233.737985&23.503187&0.01813&13.94&0.117&254.76923&149&1.709860604\\

\bf{*05189-2524}&227.779961&-30.749102&NO&79.743025&-25.412818&&&0.093&171.69231&906&0.189505861\\

\bf{NGC 5135}&311.747863&32.45034&NO&201.433583&-29.833667&0.01372&12.88&0.087&75.6&179&0.422346369\\

\bf{ARP 148}&174.18998&63.98563&NO&165.97167&40.84917&0.03452&&0.085&110.76923&914&0.121191718\\

\bf{*VIIZw31}&133.180207&22.614913&NO&79.1935&79.670167&0.05367&15.8&0.078&271.38462&835&0.325011521\\

ARP 299&141.898935&55.407582&NO&172.139559&58.563144&0.01041&11.98&0.072&58.15385&133&0.437246992\\

NGC 2146&135.653411&24.896351&NO&94.657125&78.357028&0.00298&11.38&0.071&26.58462&39&0.681656923\\

\bf{NGC 7130}&9.943251&-50.352306&NO&327.08125&-34.951306&0.01615&12.98&0.071&90.55385&197&0.459664213\\

NGC 7771&104.26359&-40.573953&NO&357.853667&20.111833&0.01427&13.08&0.07&180&84&2.142857143\\

\bf{IC 1623}&145.196075&-79.665042&NO&16.946583&-17.507028&0.02007&&0.065&235.38462&107&2.199856262\\

\bf{MRK 331}&104.461088&-40.124167&NO&357.861677&20.586075&0.01848&14.87&0.064&92.76923&228&0.406882588\\

NGC 253&97.369238&-87.964044&NO&11.888002&-25.28822&0.0008&8.04&0.059&7.47692&3&2.492306667\\

\bf{NGC 7469}&83.0985253&-45.4666436&YES&345.8150955&8.8739973&0.01632&13&0.059&60.64615&257&0.235977237\\

NGC 3893&148.155029&65.227411&NO&177.159125&48.710833&0.00323&11.16&0.056&6.36923&99&0.064335657\\

\bf{NGC 1365}&237.956197&-54.598032&NO&53.401548&-36.140402&0.00546&10.32&0.053&85.84615&15&5.723076667\\

M82&141.409718&40.566798&NO&148.96758&69.679704&0.00068&9.3&0.053&8.30769&4&2.0769225\\

NGC 6946&95.718973&11.672974&YES&308.718068&60.153946&0.00016&9.61&0.053&13.56923&7&1.938461429\\

NGC 5775&359.431609&52.423009&NO&223.489988&3.544458&0.00561&12.24&0.052&15.78462&87&0.181432414\\

\bf{NGC 1614}&204.451101&-34.381535&NO&68.499394&-8.578883&0.01594&13.63&0.051&34.61538&341&0.101511378\\

IC 342&138.172575&10.579954&YES&56.702125&68.096111&0.0001&9.1&0.05&13.01538&3&4.33846\\

NGC 5005&101.6143082&79.2490978&NO&197.7342958&37.0592056&0.00316&10.61&0.049&11.35385&50&0.227077\\

NGC 1022&179.0179945&-57.3715587&NO&39.6362718&-6.6774286&0.00485&12.09&0.047&5.53846&222&0.024948018\\

NGC 4945&305.272076&13.3398849&YES&196.3644897&-49.4682129&0.00187&9.3&0.047&7.47692&5&1.495384\\

\bf{18293-3413}&0.1481729&-11.3073324&YES&278.171375&-34.1909722&0.01818&&0.047&111.6&127&0.878740157\\

\bf{NGC 695}&140.583698&-38.230978&NO&27.809342&22.582363&0.03247&13.84&0.046&119.07692&401&0.296949925\\

NGC 4041&132.701938&54.047245&NO&180.55075&62.137278&0.00412&11.88&0.046&4.98462&174&0.028647241\\

\bf{MRK 1027}&157.758852&-52.01998&NO&33.52329&5.173238&0.03022&&0.045&52.33846&765&0.068416288\\

NGC 5678&100.044725&54.504871&NO&218.023208&57.921444&0.00641&12.13&0.044&20.76923&94&0.220949255\\

M 83&314.583573&31.972746&NO&204.253833&-29.86575&0.00172&8.2&0.043&9.69231&4&2.4230775\\

NGC 3079&157.810244&48.359895&NO&150.490827&55.679743&0.00375&11.54&0.042&27.69231&23&1.204013478\\

NGC 7479&86.270769&-42.841757&NO&346.236042&12.322889&0.00794&11.6&0.042&31.01538&97&0.319746186\\

\bf{NGC 6701}&90.396653&24.401574&NO&280.801917&60.653333&0.01323&13.01&0.041&38.21538&199&0.192037085\\

NGC 7331&93.7219374&-20.7241007&NO&339.2670653&34.4156372&0.00272&10.35&0.041&12.18462&44&0.276923182\\

NGC 520&138.705785&-58.060572&NO&21.146125&3.792417&0.00761&&0.039&17.72308&124&0.142928065\\

NGC 2276&127.669983&27.709215&NO&111.809833&85.754556&0.00804&11.93&0.039&11.07692&258&0.042933798\\

NGC 4631&142.805542&84.223494&NO&190.533375&32.5415&0.00202&9.75&0.037&2.21538&60&0.036923\\

NGC 660&141.60674&-47.346341&NO&25.759792&13.645667&0.00283&12.02&0.036&7.2&56&0.128571429\\

NGC 2903&208.711567&44.540153&NO&143.042125&21.500833&0.00186&9.68&0.036&2.49231&35&0.071208857\\

NGC 4030&277.370075&59.214755&NO&180.0985&-1.1&0.00487&12.01&0.036&14.95385&40&0.37384625\\

NGC 3628&240.851695&64.780866&NO&170.070917&13.5895&0.00281&10.28&0.034&6.64615&17&0.39095\\

NGC 4414&174.538951&83.181944&NO&186.612917&31.223528&0.00239&10.96&0.033&4.43077&39&0.113609487\\

\bf{ARP 55}&176.29697&43.9411&NO&138.97958&44.33194&0.0393&&0.03&105.23077&443&0.237541242\\

NGC 4826&315.680884&84.421337&NO&194.182333&21.681083&0.00136&9.36&0.03&1.10769&36&0.030769167\\

NGC 1055&171.331191&-51.749461&NO&40.438458&0.443167&0.00332&11.4&0.028&10.24615&34&0.301357353\\

NGC 5713&351.023352&52.123848&NO&220.048042&-0.289222&0.00658&12.18&0.027&6.09231&149&0.040887987\\

M 51&104.8513511&68.5608401&NO&202.4696292&47.1951722&0.00154&8.96&0.026&13.84615&10&1.384615\\

\bf{NGC 1144}&175.8754354&-49.8886843&NO&43.8008169&-0.1835573&0.02885&13.78&0.025&73.93846&264&0.280069924\\

NGC 891&140.383473&-17.413846&YES&35.639224&42.349146&0.00176&10.81&0.024&6.92308&20&0.346154\\

NGC 828&139.188076&-21.183497&NO&32.539875&39.190361&0.01793&13.15&0.022&36&203&0.177339901\\

NGC 1530&135.22306&17.764624&YES&65.862917&75.295583&0.00821&12.25&0.021&13.56923&114&0.119028333\\

NGC 3556&148.314482&56.251842&NO&167.879042&55.674111&0.00233&10.69&0.02&2.49231&52&0.047929038\\

NGC 3627&241.96111&64.418688&NO&170.062615&12.991549&0.00243&9.65&0.017&2.21538&27&0.082051111\\

NGC 3147&136.2899917&39.463239&NO&154.2235454&73.4007486&0.00941&11.43&0.015&24.92308&55&0.453146909\\

NGC 5055&105.996973&74.28768&NO&198.955542&42.029278&0.00168&9.31&0.012&2.76923&13&0.213017692\\

\enddata

\tablecomments{ The table shows some properties of the galaxies in
the HCN Survey. LIGs with  $L_{IR} \ge 10^{11}L_{\odot}$ are
highlighter in boldface and ULIGs with $L_{IR} \ge 10^{12}L_{\odot}$
are further indicated by *. The column ``G-Plane'' indicates if the
galaxy is near the Galactic Plane. In that case the galaxy (case
``YES'') was excluded from consideration in our analysis.}

\end{deluxetable}


\tabletypesize{\small}
\begin{deluxetable}{llccc}
\tablenum{2}
\tablecaption{The HCN survey results}
\tablehead{
 \colhead{Subclass} &\colhead{Sorting Criteria} & \colhead{Max N. \tablenotemark{a}} & \colhead{Max  $\sqrt{TS}$ \tablenotemark{b} } & \colhead{N. Y Max \tablenotemark{c} } 
}
\startdata
               
ALL                        & Distance (Mpc)              & 36  & 0.7  & 18 \\
ALL                        & $(SFR/SFR_{MW})/k$          & 54  & 0.4  & 6  \\
ALL                        & $L_{HCN}/L_{CO}$            & 54  & 1.2  & 4  \\
ALL                        & k-factor                    & 26  & 1.1  & 20 \\
ULiGs $L >10^{12} L_{sol}$ &   red shift                 & 8   & 1.8  & 6  \\
ULiGs $L >10^{12} L_{sol}$ & $(SFR/SFR_{MW})/k$          & 8   & 1.3  & 2  \\
ULiGs $L >10^{12} L_{sol}$ & $L_{HCN}/L_{CO}$            & 8   & 0.5  & 4  \\  
$L>10^{11} L_{sol}$        & $(SFR/SFR_{MW})/k$          & 24  & 0.7  & 6  \\
$L>10^{11} L_{sol}$        & $L_{HCN}/L_{CO}$            & 24  & 1.2  & 4  \\

\enddata

\tablenotetext{a}{Maximum number of objects considered.}
\tablenotetext{b}{Maximum $\sqrt{TS}$ found.}
\tablenotetext{c}{Number of objects yielding the maximum $\sqrt{TS}$.}

\end{deluxetable}


\begin{deluxetable}{lccccc}

\singlespace

\tabletypesize{\scriptsize}


\tablewidth{0pt}

\tablenum{3}

\tablecaption{Flux Upper Limits}

\tablehead{

\colhead{Name} & & \colhead{Energy range [MeV]} \\

\cline{2-6} & \colhead{100-300}  & \colhead{300-1000}  &

\colhead{$>$100}   &

\colhead{$>$300}   & \colhead{$>$1000}  \\

} \startdata

ARP 193   & 5 & 1 & 4 & 1 & 1\\

MRK 273   & 4 & 1 & 3 & 1 & 0\\

MRK 231   & 6 & 1 & 6 & 2 & 1\\

UGC 05101 & 7 & 1 & 5 & 1 & 0\\

23365+3604& 6 & 2 & 5 & 2 & 1\\

NGC  1068 & 4 & 1 & 4 & 1 & 1\\

10566+2448& 4 & 1 & 3 & 1 & 0\\

IC 5179    & 6 & 1 & 5 & 2 & 1\\

ARP 220    & 6 & 3 & 6 & 3 & 1\\

05189-2524& 8 & 2 & 9 & 3 & 2\\

NGC 5135   & 4 & 2 & 3 & 2 & 0\\

ARP 148   &  3&  1& 3 & 1 & 0\\

VIIZw31    & 4 & 1 & 3 & 1 & 1\\

ARP 299    & 3 & 1 & 2 & 1 & 0\\

NGC 2146  & 3 & 1 & 3 & 1 & 0\\

NGC 7130  & 5 & 2 & 5 & 3 & 2\\

NGC 7771  & 9 & 3 & 8 & 4 & 2\\

IC 1623   & 5 & 2 & 4 & 2 & 1\\

MRK 331   &10 & 2 & 8 & 4 & 2\\

NGC 253   & 6 & 1 & 8 & 2 & 2\\

NGC 3893  & 3 & 1 & 3 & 1 & 0\\

NGC 1365  & 5 & 1 & 5 & 2 & 1\\

M82       & 3 & 1 & 3 & 1 & 0\\

NGC 5775  & 9 & 2 & 5 & 2 & 1\\

NGC 1614   & 6 & 2 & 5 & 1 & 1\\

NGC 5005   & 5 & 1 & 4 & 1 & 1\\

NGC 1022  & 4 & 1 & 4 & 2 & 1\\

NGC 695    & 5 & 1 & 4 & 1 & 1\\

NGC 4041   & 3 & 1 & 2 & 1 & 1\\

MRK 1027   & 5 & 1 & 4 & 1 & 1\\

NGC 5678   & 4 & 1 & 3 & 1 & 1\\

M83        & 4 & 1 & 3 & 1 & 0\\

NGC 3079   & 3 & 1 & 3 & 2 & 1\\

NGC 7479   & 5 & 2 & 4 & 2 & 0\\

NGC 6701   &10 & 6 &12 & 6 & 1\\

NGC 7331   & 7 & 2 & 8 & 2 & 1\\

NGC 2276   & 4 & 3 & 5 & 4 & 1\\

NGC 520    & 7 & 2 & 7 & 2 & 1\\

NGC 4631   & 3 & 1 & 3 & 1 & 0\\

NGC 660    & 5 & 1 & 6 & 1 & 1\\

NGC 2903   & 4 & 1 & 4 & 1 & 1\\

NGC 4030   & 3 & 1 & 2 & 1 & 1\\

NGC 3628   & 3 & 2 & 5 & 2 & 0\\

NGC 4414   & 5 & 2 & 3 & 2 & 1\\

ARP 55     & 4 & 2 & 4 & 2 & 2\\

NGC 4826   & 4 & 1 & 4 & 1 & 0\\

NGC 1055   & 4 & 1 & 4 & 1 & 1\\

NGC 5713   & 5 & 1 & 4 & 1 & 0\\

M51        & 8 & 3 & 7 & 3 & 1\\

NGC 1144   & 4 & 1 & 4 & 2 & 2\\

NGC 828    & 9 & 1 & 8 & 2 & 1\\

NGC 3556   & 3 & 1 & 2 & 1 & 0\\

NGC 3627   & 3 & 2 & 5 & 2 & 0\\

NGC 3147   & 4 & 1 & 4 & 1 & 0\\

NGC 5055   & 6 & 1 & 6 & 1 & 1\\

\hline

\enddata

\tablecomments{Energy values are in MeV and  fluxes in units of
$10^{-8}$ photons cm$^{-2}$ s$^{-1}$. }

\end{deluxetable}

\end{document}